\documentclass[12pt]{spieman}  
\usepackage{amsmath,amsfonts,amssymb}
\usepackage{graphicx}
\usepackage{setspace}
\usepackage{tocloft}
\usepackage{multirow}
\usepackage{tablefootnote}
\usepackage{lineno}

\title{Starshade Exoplanet Data Challenge}

\author[a,b,*]{Renyu Hu}
\author[a,c]{Sergi R. Hildebrandt}
\author[a]{Mario Damiano}
\author[a]{Stuart Shaklan}
\author[a]{Stefan Martin}
\author[a]{Doug Lisman}

\affil[a]{Jet Propulsion Laboratory, California Institute of Technology, Pasadena, CA 91109, USA}
\affil[b]{Division of Geological and Planetary Sciences, California Institute of Technology, Pasadena, CA 91125, USA}
\affil[c]{Division of Physics, Mathematics and Astronomy, California Institute of Technology, Pasadena, CA 91125, USA}

\cftpagenumbersoff{figure}
\cftpagenumbersoff{table} 
\begin{document} 
\maketitle

\begin{abstract}
Starshade in formation flight with a space telescope is a rapidly maturing technology that would enable imaging and spectral characterization of small planets orbiting nearby stars in the not-too-distant future. While performance models of the starshade-assisted exoplanet imaging have been developed and used to design future missions, their results have not been verified from the analyses of synthetic images. Following a rich history of using community data challenges to evaluate image-processing capabilities in astronomy and exoplanet fields, the Starshade Technology Development to TRL5 (S5), a focused technology development activity managed by the NASA Exoplanet Exploration Program, is organizing and implementing a starshade exoplanet data challenge. The purpose of the data challenge is to validate the flow down of requirements from science to key instrument performance parameters and to quantify the required accuracy of noisy background calibration with synthetic images. This data challenge distinguishes itself from past efforts in the exoplanet field in that (1) it focuses on the detection and spectral characterization of small planets in the habitable zones of nearby stars, and (2) it develops synthetic images that simultaneously include multiple background noise terms -- some specific to starshade observations -- including residual starlight, solar glint, exozodiacal light, detector noise, as well as variability resulting from starshade’s motion and telescope jitter. In this paper, we provide an overview of the design and rationale of the data challenge. Working with data challenge participants, we expect to achieve improved understanding of the noise budget and background calibration in starshade-assisted exoplanet observations in the context of both Starshade Rendezvous with {\it Roman} and HabEx. This activity will thus help NASA prioritize further technology developments and prepare the science community for analyzing starshade exoplanet observations.
\end{abstract}

\keywords{Starshade, High Contrast Imaging, Exoplanet, Data Challenge, Roman Space Telescope, HabEx}

{\noindent \footnotesize\textbf{*}Further author information: e-mail:\linkable{renyu.hu@jpl.nasa.gov} \\ @2021 California Insitute of Technology. Government sponsorship acknowledged. }

\begin{spacing}{2}   

\section{Introduction}
\label{sec:intro}

Data challenges have advanced the planning and development of major astronomy facilities both on the ground and in space. Science communities participate in the data challenges to analyze simulated data and gain insight into the detection capabilities of the instrument; in turn, instrument designers learn the precision and noise level needed to reveal the objects and phenomena that are looked for. For example, a series of image analysis challenges have been carried out to develop and test methods to measure weak gravitational lensing from small distortion of galaxies's shapes (e.g., the GRavitational lEnsing Accuracy Testing (GREAT) 3\cite{2014ApJS..212....5M}). These community exercises have been instrumental in consolidating the science plans for space missions like {\it Euclid}. Another prominent example is the data challenges organized by the teams of LIGO-Virgo\cite{2015PhRvD..92f3002M} and LISA\cite{2015PhRvD..92f3002M}, which helped quantify the measurement precision needed to detect gravitational waves and inform the development of these experiments. Data challenges have also advanced many exoplanet projects. For example, a radial velocity fitting challenge was conducted to find the most efficient ways to extract planetary signals embedded in stellar noises\cite{dumusque2016radial,dumusque2017radial}. For another example, a carefully planned data challenge has helped resolve discrepancies among groups in their data reduction and analysis approaches in exoplanet transit observations with {\it Spitzer}\cite{2016AAS...22840104I}.

More recently, community data challenges have been carried out to derive capabilities and inform instrument designs for the exoplanet science with {\it Roman Space Telescope}\cite{Spergel2015,Akeson2019}. Past efforts include a challenge to efficiently identify and analyze exoplanetary microlensing events from large datasets\cite{2018AAS...23115806S}, and the {\it Roman} Exoplanet Imaging Data Challenge\cite{2018AAS...23115803H,2019AAS...23314044M}. The latter, probably the first community data challenge for space-based exoplanet imaging\footnote{\url{https://www.exoplanetdatachallenge.com}}, includes efforts to validate models of planetary reflected-light spectra at {\it Roman}'s Coronagraph Instrument (CGI)'s wavelengths, to detect planets from simulated {\it Roman} images, and to determine planetary orbits from multi-epoch observations. The {\it Roman} Exoplanet Imaging Data Challenge has also explored enhanced planet detection and orbit determination capability with a starshade rendezvous at a late phase of the mission\cite{Seager2019}, while still focusing on detecting Jupiter-sized planets. At the time of writing, the final results of the {\it Roman} Exoplanet Imaging Data Challenge have not been published; but the atmospheric modeling effort (as Phase I of that effort) succeeded in establishing the ability to retrieve key atmospheric parameters from simulated planetary spectra\cite{2018AAS...23115803H}.

Together with the development of the concepts of Starshade Rendezvous with {\it Roman}\cite{Seager2019} and {\it Habitable Exoplanet Observatory} (HabEx)\cite{Gaudi2020}, both involving a starshade, NASA’s Exoplanet Exploration Program (ExEP) is executing the Starshade Technology Development Activity to TRL5 (S5) to rapidly mature the technology and close gaps in optical performance, formation flying, and mechanical precision and stability. Together with S5, ExEP has chartered a Science and Industry Partnership (SIP) to engage the broader science and technology communities during the execution of the S5 activity. A key recommendation that emerged from SIP meetings and discussions is to produce ``a flow down of requirements from science to key performance parameters based on synthetic images (rather than scaling formulas only)'' and ``a plan for the starshade data challenge.''  Responding to the community recommendation, S5 is now organizing and implementing a Starshade Exoplanet Data Challenge.
 

The Starshade Exoplanet Data Challenge seeks to verify and improve the exoplanet yield estimates\cite{turnbull2012search,stark2016maximized,stark2019exoearth,Hu2020JATIS} by using synthetic images that realistically capture instrumental effects due to the starshade and the telescope. With the completion of most of S5's technology milestones on instrument contrast\cite{Harness2019a,Harness2019b}, solar glint\cite{Hilgemann2019}, and formation flying\cite{Flinois2018}, we can now simultaneously include in the images multiple sources of background and noise including residual starlight, solar glint, exozodiacal light, detector noise, as well as variability resulting from starshade's motion in formation flight and telescope's jitter. Many of these terms are specific to starshade observations; while some of them may be included in past exoplanet yield estimates\cite{turnbull2012search,stark2016maximized,stark2019exoearth,Hu2020JATIS}, the interplay of these terms of background and their noises can only be revealed and evaluated with the analyses of synthetic images.

A key science question that the Starshade Exoplanet Data Challenge is designed to answer is to what extent the background can be calibrated in the context of starshade-assisted exoplanet imaging. If the background is removed to its photon-noise limit, Starshade Rendezvous with {\it Roman} could provide nearly photon-limited spectroscopy of temperate and Earth-sized planets of F, G, and K stars $<4$ parsecs away, and HabEx could extend this capability to many more stars within 8 parsecs\cite{Hu2020JATIS}. To achieve these capabilities, the flux of exozodiacal light within the planet's point spread function often needs to be calibrated to a precision better than 1\% and the solar glint better than 5\%\cite{Hu2020JATIS}. The challenges for photon-limited background calibration may come from the fact that the solar glint varies with the solar angle and the starshade’s position and orientation\cite{Hilgemann2019} and that an exoplanetary dust disk is likely inclined and may have structures created by dynamical interactions with embedded planets\cite{stark2008detectability,stark2011transit}. Also, the expected use of slit-prism spectroscopy by {\it Roman} CGI may create complexity in spectral extraction together with the background. The Starshade Exoplanet Data Challenge will thus provide the opportunity to quantify the accuracy and precision of noisy background calibration for detection and spectral characterization of small exoplanets. 

We expect the outcomes of the data challenge to help NASA identify and prioritize the areas of future technology development. As we assess the abilities to extract planets, dust structures, and their spectra from images that include varying levels of instrumental effects, we will improve our understanding of how critical the instrument performance parameters are. For example, we learn from signal-to-noise ratio calculations that an instrumental contrast of $10^{-10}$ is likely not needed for many science observations with Starshade Rendezvous, while suppressing solar glint and other stray light sources is of paramount importance\cite{Hu2020JATIS}. The data challenge will tell us, with fidelity, how much performance would be lost if the contrast were $10^{-9}$ and the brightness of solar glint were a few times higher than the current best estimate (CBE).
To summarize, the Starshade Exoplanet Data Challenge is designed to validate the flow down of requirements from science to key performance parameters, quantify the required accuracy and precision of noisy background calibration, and prepare the science community for analyzing starshade exoplanet observations. 


\section{Designs and Rationale of the Starshade Exoplanet Data Challenge}
\label{sec:data}

\subsection{Overall Structure}
\label{sec:flow}

\begin{figure}[h]
\centering
\includegraphics[width=0.9\textwidth]{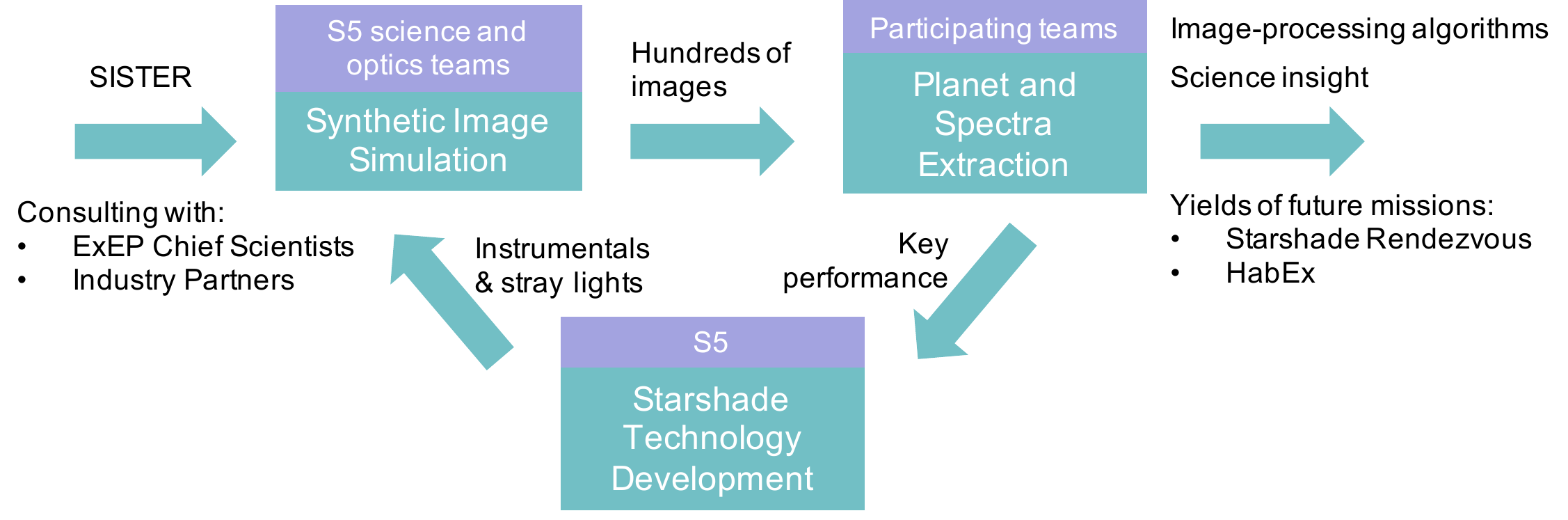}
\caption{Overall structure and workflow of the Starshade Exoplanet Data Challenge.}
\label{fig:schematic}
\end{figure}

Fig. \ref{fig:schematic} shows the overall structure and workflow of the Starshade Exoplanet Data Challenge. S5 will simulate the images for the data challenge. The images will be generated with the Starshade Imaging Simulation Toolkit for Exoplanet Reconnaissance (SISTER\cite{Hildebrandt2020JATIS}, \url{http://sister.caltech.edu}), which takes into account the full 2-dimensional nature of the astrophysical scene and the spatial variation of the Point Spread Function (PSF) due to the optical diffraction from the starshade. The simulations adopt the nominal performance parameters from current S5 results\cite{Flinois2018,Harness2019a,Harness2019b,Hilgemann2019}, including the new optical edge coating that reduces the solar glint by a factor of 10\cite{McKeithen2020JATIS}. Astrophysical and observational scenarios are selected to represent key science objectives of the well-studied starshade mission concepts including {\it Roman} Rendezvous\cite{Seager2019} and HabEx\cite{Gaudi2020} (Table \ref{tab:astro}). To explore these astrophysical and observational scenarios, as well as key instrument performance parameters (Table \ref{tab:inst}), approximately 400 images will be simulated.

Participating teams of the data challenge will then develop image-processing algorithms to test the ability to retrieve faint exoplanet signals from the synthetic images and quantify the precision of background calibration. The participating teams will attempt to determine from the images the number of the planets and their locations and brightness, as well as to extract the inclination, density, and potential structures in the exozodiacal dust disk. Estimating the uncertainties of these parameters are essential, because the resulting S/N would be compared with the S/N estimated from idealized exposure time calculators\cite{Hu2020JATIS}. With the simulated images of slit-prism spectroscopy for {\it Roman} and the data cubes of integral field spectroscopy for HabEx, the participating teams will also attempt to extract the planets’ spectra. Results from the analyses will determine the detection limit of planets vis-\`a-vis instrument parameters and indicate how well image-processing algorithms can subtract the background to the photon-noise limit. These results will inform S5 of a realistic noise budget of starshade exoplanet observations and requirements on key instrument performance parameters. The algorithms and science insight gained in the study will be disseminated among science communities via publications and code releases. 

The first set of synthetic images has been released to the public in January 2021 and the community participation of the data challenge has started\footnote{\url{https://exoplanets.nasa.gov/exep/technology/starshade-data-challenge/}}. The data challenge is scheduled to be completed by September 2021.

\subsection{Astrophysical and Observational Scenarios}
\label{sec:scenario}

\begin{table}[h]
\caption{Astrophysical and observational scenarios adopted by the Starshade Exoplanet Data Challenge. *Blue = 425-552 nm, Green = 615-850 nm (see text). 1 zodi = surface opacity of the exozodiacal dust disk in the habitable zone of the star the same as the surface opacity of the zodiacal disk at 1 AU of the Solar System\cite{Stark2014}.}
\centering
\begin{tabular}{lllll} 
 \hline\hline
 {\bf Star} & {\bf Exozodi} & {\bf Planets} & \textbf{\textit{Roman} Rendezvous} & {\bf HabEx} \\
 \hline
$\tau$ Ceti & \multirow[t]{3}{4cm}{3, 10, \& 30 zodis$^*$, Inclination 35$^{\circ}$, Smooth vs. Clumpy} & \multirow[t]{3}{3.5cm}{Two super-Earths, one hypothetical 1.0-R$_{\oplus}$ planet} & \multirow[t]{3}{3.6cm}{Imaging in blue$^*$ and green$^*$, spectroscopy in green} & \multirow[t]{3}{2.5cm}{Spectroscopy 0.3--1.0 $\mu$m \& 1.0--1.8 $\mu$m} \\
&&&&\\
&&&&\\
\hline
$\epsilon$ Indi A & \multirow[t]{3}{4cm}{1, 3, \& 10 zodis, Inclination 30$^{\circ}$ \& 80$^{\circ}$, Smooth vs. Clumpy} & \multirow[t]{3}{3.5cm}{Multiple hypothetical 1.0-R$_{\oplus}$ planets} & \multirow[t]{3}{3.6cm}{Imaging in blue and green, Spectroscopy in green} & \multirow[t]{3}{2.5cm}{Spectroscopy 0.3--1.0 $\mu$m \& 1.0--1.8 $\mu$m} \\
&&&&\\
&&&&\\
\hline
$\sigma$ Draconis & \multirow[t]{3}{4cm}{1, 3, \& 10 zodis, Inclination 30$^{\circ}$ \& 80$^{\circ}$, Smooth vs. Clumpy} & \multirow[t]{3}{3.5cm}{Multiple hypothetical 1.6-R$_{\oplus}$ planets} & \multirow[t]{3}{3.6cm}{Imaging in blue and green, Spectroscopy in green} & N/A \\
&&&&\\
&&&&\\
\hline
$\sigma$ Draconis & \multirow[t]{3}{4cm}{1, 3, \& 10 zodis, Inclination 30$^{\circ}$ \& 80$^{\circ}$, Smooth vs. Clumpy} & \multirow[t]{3}{3.5cm}{Multiple hypothetical 2.4-R$_{\oplus}$ planets} & \multirow[t]{3}{3.6cm}{Imaging in blue and green, Spectroscopy in green} & N/A \\
&&&&\\
&&&&\\
\hline
$\sigma$ Draconis & \multirow[t]{3}{4cm}{1, 3, \& 10 zodis, Inclination 30$^{\circ}$ \& 80$^{\circ}$, Smooth vs. Clumpy} & \multirow[t]{3}{3.5cm}{Multiple hypothetical 1.0-R$_{\oplus}$ planets} & N/A & \multirow[t]{3}{2.5cm}{Spectroscopy 0.3--1.0 $\mu$m \& 1.0--1.8 $\mu$m} \\
&&&&\\
&&&&\\
\hline
$\beta$ CVn & \multirow[t]{3}{4cm}{1, 3, \& 10 zodis, Inclination 30$^{\circ}$ \& 80$^{\circ}$, Smooth vs. Clumpy} & \multirow[t]{3}{3.5cm}{Multiple hypothetical 2.4-R$_{\oplus}$ planets} & \multirow[t]{3}{3.6cm}{Imaging in blue and green, Spectroscopy in green} & N/A \\
&&&&\\
&&&&\\
\hline
$\beta$ CVn & \multirow[t]{3}{4cm}{1, 3, \& 10 zodis, Inclination 30$^{\circ}$ \& 80$^{\circ}$, Smooth vs. Clumpy} & \multirow[t]{3}{3.5cm}{Multiple hypothetical 1.0-R$_{\oplus}$ planets} & N/A & \multirow[t]{3}{2.5cm}{Spectroscopy 0.3--1.0 $\mu$m \& 1.0--1.8 $\mu$m} \\
&&&&\\
&&&&\\
 \hline\hline
\end{tabular}
\label{tab:astro}
\end{table}

Table \ref{tab:astro} lists the astrophysical and observational scenarios adopted as the representative cases for the data challenge. These scenarios are chosen to probe the key and limiting science objectives of Starshade Rendezvous with {\it Roman} and HabEx. Four stars will be included in the study, including two stars $<4$ parsecs away ($\tau$~Ceti and $\epsilon$~Indi~A), one star in the 5 -- 6-parsec distance range ($\sigma$~Draconis), and one star in the $\sim8$-parsec distance range ($\beta$~CVn). These stars are in the nominal target lists of both Starshade Rendezvous with {\it Roman} and HabEx\cite{Seager2019,Gaudi2020}.

While detecting and spectrally characterizing Earth-sized planets in the habitable zones of all these stars belongs to HabEx's science goals, this would only be possible for $<4$-parsec stars with {\it Roman}\cite{Hu2020JATIS,RomeroWolf2020JATIS}. Therefore, we assume hypothetical 1.0-$R_{\oplus}$ planets when simulating HabEx observations, and larger planets when simulating {\it Roman} observing $\sigma$~Draconis and $\beta$~CVn. For the larger planets, we consider the two dominant populations of planets discovered by {\it Kepler}\cite{fulton2018california}: the ``super-Earth'' population with a representative radius of 1.6 $R_{\oplus}$ and a larger-radius population with a representative radius of 2.4 $R_{\oplus}$. The super-Earth populations are likely dominated by large rocky planets\cite{rogers2015most}, and the larger-radius population can either be rocky planets with massive H$_2$/He gas envelopes\cite{owen2017evaporation,jin2018compositional} or planets with massive water envelopes\cite{zeng2019growth,mousis2020irradiated}. We use Exo-REL\cite{Hu2019B2019ApJ...887..166H}, a well-documented model for exoplanet clouds and reflected-light spectra, to simulate the input spectra of the planets.

The star $\tau$~Ceti deserves special attention because of its known planets and outer dust disk. The analyses of radial-velocity data have not reached a consensus, but the two outer planets (the planets e and f) near its habitable zone are consistent between analyses\cite{tuomi2013signals,feng2017color}. A debris disk has been detected with far-infrared and radio observations \cite{greaves2004debris,lawler2014debris,macgregor2016alma}, with an inner edge at $\sim6$ AU , an outer edge at $\sim50$ AU, and an inclination of $\sim35^{\circ}$. The knowledge motivates us to consider the following in the data challenge for $\tau$~Ceti. (1) We adopt the debris disk's inclination as the inclination of the planets and the exozodiacal disk. As such, the true masses of the planets e and f are $\sim6.9$ $M_{\oplus}$. We further consider the possibility that the planets are either predominantly rocky or with large water envelopes\cite{zeng2019growth} in estimating their radii and simulating the spectra. (2) We include the possibility of a denser exozodiacal disk to test its impact on planet detection (Table \ref{tab:astro}). The LBTI exozodiacal disk survey did not detect a disk at $\tau$ Ceti, but the $1-\sigma$ upper limit is 44 zodis\cite{Ertel2020}. (3) We include another planet, with Earth's radius and Earth's mass, between the orbits of the planets e and f. This hypothetical planet is predicted by orbital dynamics and exoplanet population-level information\cite{dietrich2020integrated}. We verify that the planet would be dynamically stable\cite{kane2016catalog} and would induce a radial-velocity signal amplitude of $\sim6$ cm s$^{-1}$, which is well below the detection limit of existing data.

In this data challenge, we will test the ability to detect planets embedded in their exozodiacal disk. Because it is not practical to simulate a self-consistent disk with the planets assumed -- given the uncertainties in the source of the particles, the existence of outer planets, as well as the particle size distributions -- we instead attempt to bound the problem by considering the endmembers of a ``smooth'' disk and a ``clumpy'' disk. For the ``smooth'' disk we adopt a solar-system disk density profile and use Zodipic\cite{kuchner2012zodipic} to simulate the intensity including the effects of inclination and particle forward scattering\cite{stark2011transit}. The ``clumpy'' disk represents a more challenging scenario for planet detection, where clumps of dust particles are trapped in mean-motion resonance with the planets\cite{stark2008detectability,stark2011transit}. To our knowledge, this would be the first time that potential structures of the exozodiacal disk are assessed against the imaging and detection of small exoplanets of nearby stars. We envision the data challenge would eventually inform us to what extent a clumpy disk would hinder the revelation of the planets in the system.

Lastly, we design the observational scenarios to mimic the basic ideas of operation outlined by Starshade Rendezvous with {\it Roman}\cite{Seager2019} and HabEx\cite{Gaudi2020}. Starshade Rendezvous with {\it Roman} would perform broadband searches of small planets, and if any feasible planets are detected, conduct spectroscopy immediately following imaging in the green band (615-800 nm)\cite{RomeroWolf2020JATIS}. We adopt this philosophy by considering two visits per astrophysical scenario, each with both broadband imaging and slit-prism spectroscopy. We will additionally explore two aspects for Starshade Rendezvous with {\it Roman}. (1) We will include broadband imaging in the blue band (425-552 nm) and compare it with the green band. The blue band would have less exozodiacal light per resolution element and may thus lead to better planet detection. (2) We will include the green-band spectroscopy up to 850 nm, where the expected instrument contrast degrades from $10^{-10}$ at 800 nm to $10^{-9}$ at 850 nm\cite{Seager2019}. This is motivated by the desire to extend coverage to longer wavelengths to reduce spectral degeneracies\cite{damiano2020multi} and the realization that an instrument contrast of $10^{-9}$ may be sufficient\cite{Hu2020JATIS}. HabEx would conduct most planet searches with its coronagraph and perform integral field spectroscopy with its starshade\cite{Gaudi2020}. We thus focus on spectroscopy for HabEx, also considering two visits per astrophysical scenario. As all scenarios adopted contain planets that fit the definition of ``high-interest'' by HabEx, we will include near-infrared spectroscopy (1.0--1.8 $\mu$m) in addition to the 0.3--1.0 $\mu$m band for all scenarios. Note that this entails two spectral integrations per visit, as the starshade needs to be located at a different separation from the telescope for conducting the near-infrared spectroscopy. Because this data challenge focuses on image processing and background calibration, it does not address potential synergy between a starshade and the CGI of \textit{Roman}, or other operational and logistical constraints of the starshade.

\subsection{Instrument Effects}
\label{sec:instrument}

\begin{table}[h]
\caption{Instrumental effects and other backgrounds explored by the Starshade Exoplanet Data Challenge.}
\centering
\begin{tabular}{lll} 
 \hline\hline
 {\bf Item} & {\bf Description} & {\bf Variation} \\
 \hline
Residual starlight & \multirow[t]{2}{8cm}{Instrument contrast of $10^{-10}$ produced by random deviation of the edge of the petals} &  \multirow[t]{2}{4cm}{Contrast level up to $1\times10^{-9}$} \\
&&\\
 \hline
Lateral displacement & \multirow[t]{2}{8cm}{Time-dependent lateral shift up to 1~m in formation flight} &  \multirow[t]{2}{4cm}{N/A} \\
&&\\
 \hline
Solar glint & \multirow[t]{2}{8cm}{Time-dependent sunlight scattered by the optical edge with coating} &  \multirow[t]{2}{4cm}{Up to 3-times brighter than the CBE} \\
&&\\
 \hline
Other stray light & \multirow[t]{2}{8cm}{Reflection of Milky Way, Earth, Jupiter, and leakage through micrometeoroid holes} &  \multirow[t]{2}{4cm}{N/A} \\
&&\\
 \hline
Local zodical light & \multirow[t]{2}{8cm}{V band magnitude of 22.5 per arcsec$^2$} &  \multirow[t]{2}{4cm}{N/A} \\
&&\\
 \hline
Telescope jitter & \multirow[t]{2}{8cm}{Random pointing error of 14 mas for {\it Roman} and 0.3 mas for HabEx} &  N/A \\
&&\\
 \hline
Integration time & \multirow[t]{2}{8cm}{Estimated for S/N per band or spectral element assuming photon-noise background calibration} &  \multirow[t]{2}{4cm}{S/N of 5, 10, and 20} \\
&&\\
 \hline\hline
\end{tabular}
\label{tab:inst}
\end{table}

Table \ref{tab:inst} lists the instrumental and other effects to be included in the synthetic images and explored by the data challenge. We will adopt the current best estimates for the residual starlight, solar glint, other stray light, and formation flying performance as summarized in Hu et al. (2021)\cite{Hu2020JATIS}. As the residual starlight contrast is the fundamental requirement that controls mechanical precision tolerance, we will vary the contrast level and see how much we could tolerate without adversely impacting planet detection and background calibration. We will adopt the latest estimate of the brightness of the solar glint\cite{McKeithen2020JATIS} and also investigate how much brighter we could tolerate. Other stray light sources such as the reflected light of the Milky Way, Earth, Jupiter, as well as the leakage through micrometeoroid holes, will be dimmer than the solar glint by a factor of at least 2\cite{Hu2020JATIS} and their impact will be studied together with the enhancement factor applied to the solar glint (Table \ref{tab:inst}).

A new effect that will be simulated for the data challenge is the time-dependent variability of these backgrounds. As the solar glint is a starshade's closest analog to speckles in coronagraphic imaging, it is important to include the temporal variability of the stray light to achieve high-fidelity image simulations. We consider telescope jitter, lateral motion of the starshade in formation flight, as well as the change in Sun's angle during long integration as the main sources of temporal variability. For instance, images will be produced with realistic random pointing errors. This way, the data challenge will be able to assess the precision of background calibration achievable from real observations and how it compares with the photon-noise limit.

In addition to broadband imaging, we will simulate spectroscopy with {\it Roman}'s slit-prism spectrometer and HabEx's integral field spectrometer. The simulations will mimic the dispersion of the planets and background sources and the recording of the spectra on the detector. The spectral resolution, as well as the width and orientation of the slit when applicable, will follow the specifications of the mission concepts. For \textit{Roman}, we will study whether a specific slit orientation would be necessary or preferred for extracting planetary spectra from interfering backgrounds such as exozodiacal light. For instance, the slit may be oriented radially along the axis between the star and the planet, perpendicular to that axis, or at some angle to both -- this may affect the ability to subtract off the exozodiacal background. To our knowledge, it will be the first time that slit-prism spectroscopy will be simulated and studied in detail for exoplanet direct imaging using a starshade. The knowledge gained in this analysis will help us better understand the background calibration and planet signal extraction in this form of spectroscopy.

Besides the parameters listed in Table \ref{tab:inst}, we will adopt the telescope and instrument parameters designed for {\it Roman} and HabEx, including the optical throughput and detector properties. The frame rate will be chosen for each observation consistent with the photon counting mode of the EMCCDs and the detector noise as the combination of dark current and clock-induced charge will be included accordingly\cite{stark2019exoearth}.

\section{Summary and Expected Outcomes}
\label{sec:conclusion}

To summarize, the Starshade Exoplanet Data Challenge will assess the noise budget of exoplanet observations using a starshade and determine the precision of background calibration achievable with synthetic images. As described in Section~\ref{sec:data}, the data challenge will be based on the hitherto most realistic simulations of starshade-assisted observations that explore the expected diversity of planet types as well as the density, structure, and inclination of the dust disks around the nearby stars. In addition to an ensemble of instrumental effects, we will include temporal variability of residual starlight and stray light due to the starshade’s motion and telescope's jitter. The high-fidelity synthetic images will help us better understand the detection limit of planets and their spectra as a function of instrument performance parameters.

Specific outcomes that may be anticipated from the data challenge include (1) estimation of the S/N of planetary parameters (e.g., location, brightness, spectrum) based on the synthetic images. The estimated S/N, in comparison with the idealized S/N used to set the integration time (Table~\ref{tab:inst}), will tell us how precisely the exozodiacal light and solar glint can be calibrated for planet detection. (2) Detection of exozodiacal disks and constraints of their density, inclination, and possible clumpy structures. We will assess what we could learn about exozodiacal dust disks from direct-imaging observations using a starshade, and also evaluate whether the clumpy structures would interfere with planet detection. (3) Extraction of planetary spectra, especially from slit-prism spectroscopy. Using synthetic images, we will evaluate whether it would be feasible to extract planetary spectra when both the planet and the extended background of exozodiacal light are dispersed onto the detector. As a whole, these insights will tell us what we could realistically expect to learn about planets and disks around nearby stars using a starshade, and how these capabilities would depend on the instrument contrast and the suppression of solar glint and other straylight.

The image-processing algorithms and science insight gained in this exercise may also advance high-contrast imaging astronomy in general. Compared to coronagraph direct imaging from the ground and in space, where creative algorithms have been developed to subtract backgrounds and speckles, starshade-assisted direct imaging mainly presents a different set of problems, associated with distinguishing planets from the exozodiacal disk, as well as removing stray light terms specific to the starshade. The data challenge will enable a cross-disciplinary development of techniques to be used to maximize the science yield of the future missions using starshades, ranging from step-by-step (e.g., image subtractions and feature extraction) to more holistic (e.g., Bayesian inversion, deep learning) image analyzing techniques. With the synthetic images, algorithms, science and technology insight, and community partnership, the Starshade Exoplanet Data Challenge will result in an enduring legacy that advances exoplanet astronomy in the years to come.

\acknowledgments     
We thank Karl Stapelfeldt, Eric Mamajek, Christopher Stark, and Phil Willems for helpful discussions. The research was carried out at the Jet Propulsion Laboratory, California Institute of Technology, under a contract with the National Aeronautics and Space Administration (80NM0018D0004).



\section*{Author Biography}

{\bf Renyu Hu} is a scientist at the Jet Propulsion Laboratory and his research strives to identify and characterize habitable environments in the Solar System and beyond. He is the Starshade Scientist of the NASA Exoplanet Exploration Program, providing science leadership to the S5 Starshade Technology Development Activity and managing the Starshade Science and Industry Partnership program. He received his PhD in Planetary Science at Massachusetts Institute of Technology in 2013.

\noindent{\bf Sergi R. Hildebrandt} is a scientist at the Jet Propulsion Laboratory and Lecturer at the California Institute of Technology. He received a PhD in Theoretical Physics at the University of Barcelona. He has worked in the data analysis of the Cosmic Microwave Background, adaptive optics in the visible, and more recently the development of SISTER, a user friendly, open source, tool that generates starshade simulations with high fidelity. 

\noindent{\bf Mario Damiano} is a Jet Propulsion Laboratory postdoc fellow. He performs data analysis to study and interpret the spectroscopic characteristics of exoplanetary atmospheres to unveil atmospheric composition and dynamics of these alien worlds. He is also an enthusiast of deep learning and AI. He received his PhD in Physics and Astronomy at the University College London in 2019.

\noindent{\bf Stuart Shaklan} is the supervisor of the High Contrast Imaging Group in the Optics Section of the Jet Propulsion Laboratory. He received his Ph.D. in Optics at the University of Arizona in 1989 and has been with JPL since 1991.

\noindent{\bf Stefan Martin} is a senior optical engineer at the Jet Propulsion Laboratory. He received his BSc degree in physics from the University of Bristol, United Kingdom, and his PhD in engineering from the University of Wales. At JPL, he has been leader of the TPF-I Flight Instrument Engineering Team, testbed lead for the TPF-I Planet Detection Testbed, and payload lead for the HabEx Telescope design study. He is currently involved in starshade accommodation on future space telescopes, such as NGRST.

\noindent{\bf Doug Lisman} is the systems engineering lead for starshade technology development at the Jet Propulsion Laboratory where he is a member of the Instrument Systems Engineering group. He received his BS degree in mechanical engineering from Washington University in St. Louis in 1984 and has been at JPL since 1984.


\end{spacing}
\end{document}